\documentstyle[aps,twocolumn,epsfig]{revtex}
\begin{document}
\draft
\title{Weak localisation in AlGaAs/GaAs $p$-type quantum wells}
\author{S. Pedersen, C.B. S\o rensen, A. Kristensen and P.E. Lindelof}
\address{The Niels Bohr Institute, University of Copenhagen, Universitetsparken 5, DK-2100 Copenhagen, Denmark}
\author{ L.E. Golub and N.S. Averkiev}
\address{ A.F. Ioffe Physico-Technical Institute, Russian Academy of Sciences,
194021 St. Petersburg, Russia.}
\date{\today}
\maketitle
\begin{abstract}
We have for the first time experimentally investigated the weak
localisation magnetoresistance in a AlGaAs/GaAs $p$-type quantum well.
The peculiarity of such systems is that spin-orbit interaction is strong.
On the theoretical side it is not possible to treat the spin-orbit
interaction as a perturbation. This is in contrast to all prior
investigations of weak localisation. In this letter we compare the experimental results with a
newly developed diffusion theory, which explicitly describes the
weak localisation regime when the spin-orbit coupling is strong. The spin relaxation rates calculated from the fitting parameters was found to agree with theoretical expectations. Furthermore the fitting parameters indicate an enhanced phase breaking rate compared to
theoretical predictions.

\end{abstract}

% body of paper here
The effect of localisation in weakly disordered systems can be
understood in terms of the quantum interference between two waves
propagating by multiple scattering along the same path but in opposite directions. When a
magnetic field is applied the phase pick up along the two paths have
opposite sign, and as a consequence, a negative magnetoresistance is observed~\cite{khel}.
This effect is normally known as {\em weak localisation}.

Due to the properties of the spin part of the wavefunction, spin-orbit interaction has been shown to have a dramatic influence on the weak localisation.
In systems with strong spin-orbit interaction the magnetoresistance reverse the sign.
This is in contrast to the above known as {\em weak antilocalisation}.

Traditionally, weak antilocalisation has been studied intensely in
metallic films \cite{hln,film}, where spin-orbit interaction occurs at the individual scattering centers. More recently weak antilocalisation has been observed in true two
dimensional systems which lack inversion symmetry, like $n$-type GaAlAs/GaAs or Te quantum wells.
The lack of inversion symmetry  gives rise to a new spin relaxation mechanism.
This has surprisingly led to a completely new physical insight
\cite{dressel,knap,110,Savelev,wurtzite,wells} (see also references in \cite{knap}).

However most of all previous works referred to $n$-type quantum wells. In
the case of a $p$-type quantum well, an even more dominating positive magnetoresistance would
be expected due to strong spin-orbit interaction in the GaAs valence band~\cite{film}.

In recent theoretical works devoted to weak localisation in $p$-quantum
wells~\cite{JETP} it was shown how the sign of the magnetoresistance
depends on the hole concentration. 
Moreover anisotropy of the spin relaxation was predicted, which in turn leads to
dependence of the phase relaxation rate on the spin orientation.
Experimental investigations of
anomalous magnetoresistance in $p$-quantum wells so far did not exist. In
this work, for the first time, the magnetoresistance is studied
experimentally in $p$-quantum wells and peculiarities of weak
localisation are discussed in the case where spin and momentum relaxation
rates are comparable.

The heterostructures used in the experiment were grown on a [100] oriented GaAs wafer by Molecular Beam Epitaxy
(MBE) technique. A symmetrical quantum well was formed as a 70\AA~wide
GaAs channel in a modulation doped Ga$_{0.5}$Al$_{0.5}$As matrix. The GaAlAs was
homogeneously doped with Be (${\rm n_{Be}}=2 \cdot 10^{18} {\rm
cm}^{-3}$) in two 50\AA~thick layers separated by 250\AA~of intrinsic
Ga$_{0.5}$Al$_{0.5}$As from the centre of the GaAs channel. The individual
samples were mesa-etched into rectangular Hall bars with a width of
0.2mm and a total length of 4.2mm. Three voltage contacts on each side
were placed in a distance of each 0.8mm to avoid perturbing significantly
the four point measurements. Ohmic contacts to the 2-dimensional hole
gas were made by a Au/Zn/Au composite film annealed at 460$^{\circ}$C in
3 minutes. The contacts areas were $0.6 \times 0.6 {\rm mm}^{2}$
squares, and bonded to the legs of a nonmagnetic chip carrier.
Four point measurements of the resistivity were carried out using
standard low frequency lock-in technique (EG$\&$G 5210). The samples
were biased by an AC current signal with an amplitude of 200nA. The
experiments were performed at temperatures between 0.3 and 1.0K in an
Oxford Heliox cryostat equipped with a copper electromagnet.
The characterisation of the samples with respect to density and mobility
were done by Hall measurement at magnetic fields between -0.3T and 0.3T,
while the weak localisation magnetoresistance measurements were
performed at fields between -100Gs and 100Gs. To generate the stable
current for the magnetic fields we used a Keithley 2400.
The samples were found to have a hole density of $p = 4.4 \cdot
10^{15}{\rm m^{-2}}$, which is low enough to ensure that only one
subband is filled. The mobility was found to be $\mu=3.5 {\rm T^{-1}}$.

\begin{figure}
\begin{center}
\epsfig{file=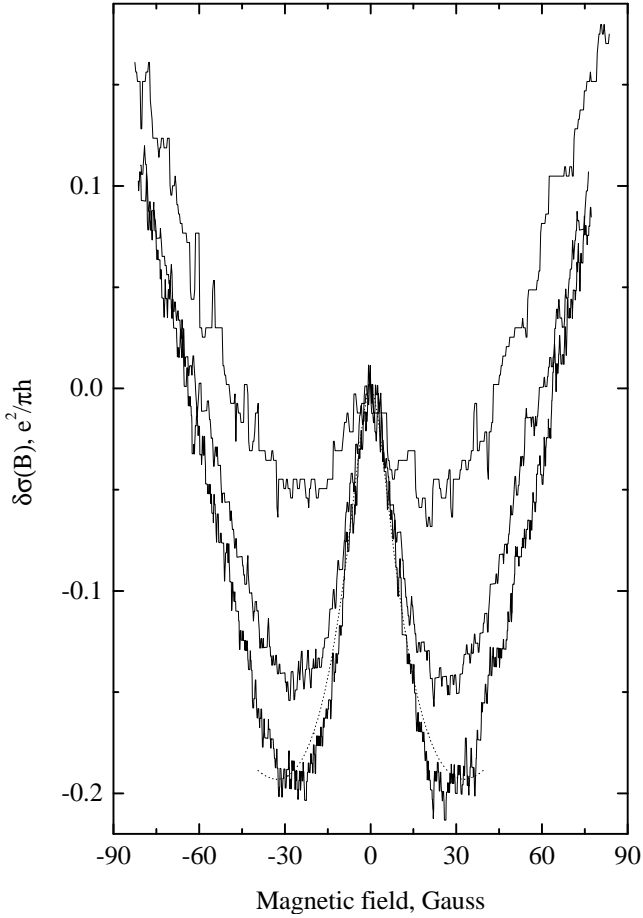,width=0.5\textwidth}
\end{center}
\caption[Til LOF]{\small{ Magnetoconductivity $\delta\sigma(B)$ of a GaAs/GaAlAs $p$-type quantum well three different temperatures (from above $\rm{T=820~mK, T=560~mK~and~ T=360~mK}$). The best theoretical fit (dotted line) is shown for $\rm{T=360~mK}$.} \protect}
\label{data}
\end{figure}

It is well known that the weak localization effect on the magnetoconductivity
manifests itself more brightly when $k_{\rm F} l \gg 1$, corresponding to a
metallic conductivity in the system. Here
$k_{\rm F}$ is the Fermi wave vector and $l$ is the mean free path.
For our samples this product may be estimated with the help of the two-dimensional Drude
conductivity, $\sigma_D$
\begin{equation}
\sigma_D = \frac{e^2}{2 \pi \hbar} \: k_{\rm F} l \:.
\end{equation}
In the studied samples $\sigma_D = 2.47 \cdot 10^{-3}~\Omega^{-1}$ which
gives $k_F l \approx 63$. The value of $k_F$ may be determined from the
hole concentration: $k_F = \sqrt{2 \pi p}$ and is equal
to $1.7 \cdot 10^8$~m$^{-1}$. This leads to a mean free path $l = 0.37$~$\mu$m for our
samples. The magnetic length is equal to $l$ in a field
$B_{tr} = \hbar / 2 e l^2 \approx 24$~Gs. For $B < B_{tr}$
the diffusion theory may be applied for description of weak localization effects.

According to recent theoretical works~\cite{JETP}, the key parameter in a
$p$-quantum well of width $a$ is $k_{\rm F} a / \pi$. This product is a measure of
heavy-hole/light-hole mixing degree at the Fermi level which determines the behavior of the anomalous magnetoresistance.
For instance, if the carrier concentration is small ($k_{\rm F} a / \pi
\ll 1$) the magnetoresistance does not change its
sign and is exclusively negative. On the other hand if $k_{\rm F} a / \pi \geq 1$, the magnetoresistance is also sign-constant, but positive. This positive magnetoresistance was observed in recent experimentally reports~\cite{ole}. 
Moreover the resistance may change its sign as a function of magnetic
field at the intermediate values of this parameter. Since in the studied
system $k_{\rm F} a / \pi \approx 0.37$ this intermediate
regime is in fact realised in our experiments.

Under these conditions the weak localisation
correction to the conductivity of our $p$-type quantum wells in magnetic fields
$B < B_{tr}$ is given as~\cite{JETP}
\begin{eqnarray}
 \label{golub}
&{}&\delta\sigma(B) = \nonumber \\ &{}&\frac{e^{2}}{\pi h}
\left[f\left(\frac{B}{B_{\varphi}+B_{\parallel}} \right)
+\frac{1}{2}
f\left(\frac{B}{B_{\varphi}+B_{\perp}} \right) - \frac{1}{2}
f\left(\frac{B}{B_{\varphi}} \right) \right], \nonumber \\ &{}& \end{eqnarray} where $f$
is given by: $f(x)=\ln(x) + \psi(1/2 + 1/x)$, here $\psi(x)$ is a
Digamma-function and
$\delta\sigma(B)$ is the difference between the conductivity with
and without magnetic field. The characteristic magnetic fields
$B_{\varphi}, B_{\parallel}$ and $B_{\perp}$ are given as
\begin{equation} \label{mag}
B_{\varphi}=\frac{\hbar}{4eD\tau_{\varphi}}, ~~~~
B_{\parallel}=\frac{\hbar}{4eD\tau_{\parallel}},
~~~~B_{\perp}=\frac{\hbar}{4eD\tau_{\perp}}, \end{equation} where the
quantities $\tau_{\parallel}$, $\tau_{ \perp}$ refer to the longitudinal
and transverse spin relaxation time with the preferred axis lying
normal to the quantum well, and $\tau_{\varphi}$ is the phase relaxation
time for the holes. The diffusion coefficient $D=l^{2} / 2 \tau$,
where $\tau$ is the momentum relaxation time. Equation~(\ref{golub}) resembles the expression
for metallic films first reported by Hikami et al.~\cite{hln} as well as that by Altschuler et al.~\cite{film} for diffusive spin-orbit effects in two-dimensional electron systems. However in our case the spin relaxation cannot be described
by one parameter and the expression given by Eq.~(\ref{golub}) does only converge into the Hikami expression
if $B_{\perp}=2B_{\parallel}$ which as we shall see is not the case.

In Fig.~\ref{data} we present the magnetoconductivity measurements at
different temperatures.
An example of a fit obtained with Eq.~(\ref{golub}) is also shown for $T=360$~mK.
The fitting was done by the Levenberg-Marquardt
method, implemented in $C^{++}$ by standard nonlinear least-squares
routines.
The parameters of the fitting procedure are:
$B_{\varphi} = 2.6$~Gs, $B_{\parallel} = 17.2$~Gs, and $B_{\perp} = 4.6$~Gs.

We have shown theoretically~\cite{JETP} that spin flip
probabilities depend differently on the value of Fermi quasimomentum
for hole spin oriented along the grown axis and lying in the quantum well
plane. For instance, for scattering  from the short-range
potential $B_\| \sim k_{\rm F}^4$ and $B_\perp \sim k_{\rm F}^6$. This
leads at arbitrary small hole concentrations to the inequality
$B_\| > B_\perp$ which is observed in the experiment.

\begin{figure}
\begin{center}
\epsfig{file=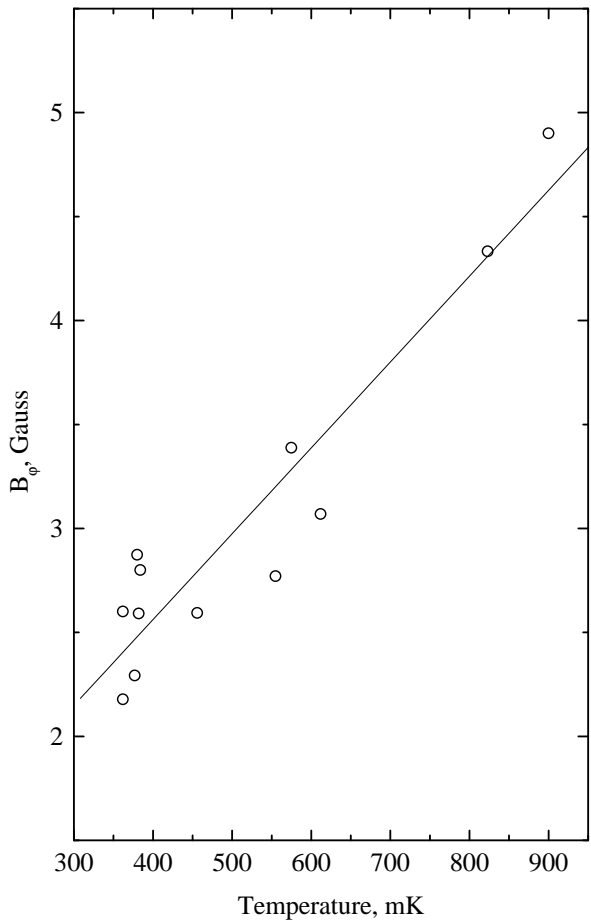,width=0.5\textwidth}
\end{center}
\caption[Til LOF]{\small{Temperature dependence of the parameter $\rm{B_{\varphi}}$ of a  GaAs/GaAlAs $p$-type quantum well.  Experimental results are shown by open dots. The solid line shows the best linear fit to the data points.} \protect}\label{final}
\end{figure}

Since $B_\varphi < B_{tr}$, the wave function phase breaks after
many collisions with impurities and one can apply the diffusion theory
for experiment fitting.
In magnetic fields $B \sim B_{tr}$  the wave function phase breaks
after a few collisions. Weak localisation theory
for this region of fields is derived in
references~\cite{Kawabata,Dmitriev} for systems with weak spin-orbit
interaction only. Below we consider the case of strong
spin-orbit interaction in magnetic fields $B \sim B_{tr}$.

The Cooperon equations for particles with different absolute
value of spin projection can be separated at $B \geq B_{tr}$~\cite{JETP}. Thus
the expression for $\delta \sigma$ has three terms and each of them depends
only on one characteristic magnetic field, similar to Eq.~(\ref{golub}). The Cooperon equations, which take
into account strong spin-orbit interaction are complicated integral
equations and have to be solved numerically. However it is clear that the
absolute value of each term in the expression for
$\delta \sigma$ decreases in comparison with the diffusion approximation.
Hence Eq.~(\ref{golub}) describes qualitatively
the dependence $\delta \sigma (B)$ even at $B \geq B_{tr}$.
The maxima and the subsequent  decrease in magnetoconductivity seen in Fig.~\ref{data}
is in fact caused by the first term in Eq.~(\ref{golub}) which dominates
in these fields.

Thus the magnetoconductivity dependence in small magnetic fields is approximately given by
\begin{equation}
 \label{smallB}
\delta\sigma(B) = - \frac{e^{2}}{48 \pi h}
\left(\frac{B}{\tilde{B}} \right)^2 \:.
\end{equation}
One can show that $\tilde{B} \approx B_\varphi$ if $B_\|, B_\perp >
B_\varphi$. At $T=360$~mK this inequality is valid.
The spin relaxation times, $\tau_\|$ and $\tau_\perp$, are temperature
independent because the studied system is degenerate and charge transport
is realised by the carriers near the Fermi surface. 

The temperature dependence of $B_{\varphi}$ is shown in Fig.~\ref{final}.
One can see that it is roughly linear.
A least square fit gives the approximation:
$B_{\varphi}(T)=4.1~{\rm Gs~K^{-1} T}+0.91~{\rm Gs}$. As an estimate
we use the Nyquist noise formula for the electron phase breaking time as
an approximation for $B_{\varphi}$, \cite{electrons}: \begin{equation}
B_{\rm N}=B_{tr} \frac{k_{\rm B}T}{\hbar k_{\rm F} v_{\rm F}} \ln \left( k_{\rm F} l \right), \end{equation}
where $v_{\rm F}$  is the hole velocity at the Fermi surface.
It is related to the mean free path by equality $l= v_{\rm F} \tau$.
In this approximation $B_{\rm N}=0.9~{\rm Gs~K^{-1} T}$, where an effective hole mass $m_{h}=0.23 \cdot
m_{0}$ was used ($m_{0}$ is the free electron mass).
Hence the observed phase breaking rate is approximately four times larger than what is
expected from this simple Nyquist noise estimate. A possible explanation
for this discrepancy could be found in the non-parabolic dispersion
relation which would tend to decrease $v_{\rm F}$. It is however difficult to
make any further analyse due to the fact there has been no theoretical
attempts to discuss the phase breaking rate in hole systems.

In conclusion, we have for the first time presented experimental studies  of the
magnetoconductivity caused by weak localisation in GaAlAs/GaAs $p$-type
quantum well system, where the spin-orbit coupling is strong. We observe that the magnetoconductance changes sign from negative to positive as the magnetic field is increased. This is due to the intermediate degree of heavy-hole/light-hole mixing in these samples. The phase
relaxation times were determined as a function of
temperature.  The spin relaxation rates are found to be in agreement
with theory. The phase coherence relaxation rate was found to be
significantly larger than the Nyquist behaviour previously
found to explain the values for electron systems.

L.E.G. and N.S.A. thanks
RFBR (grant 98-02-18424), program
``Physics of Solid State Nanostructures'' (grant 97-1035)
and Volkswagen Foundation for financial support.

The experimental part of our research was supported by Velux Fonden, Ib Henriksen Foundation, Novo Nordisk Foundation, Danish Research Council (grant 9502937, 9601677 and 9800243).

% now the references. delete or change fake bibitem. delete next three
%   lines and directly read in your .bbl file if you use bibtex.

% \begin{figure}
% \caption{}
% \label{}
% \end{figure}

\end{document}